\documentclass[prb,superscriptaddress,twocolumn,longbibliography]{revtex4-2}

\usepackage{amsmath,amssymb}
\usepackage{physics}
\usepackage{graphicx}
\usepackage{bm}
\usepackage{bbm}
\usepackage[colorlinks,bookmarks=false,citecolor=blue,linkcolor=black,urlcolor=red]{hyperref}
\usepackage{color} 
\usepackage{newtxtext,newtxmath}

\allowdisplaybreaks

\begin{document}

\title{
Neural decoders for subsystem many-hypercube codes
}

\date{\today}

\author{Ryota Nakai}
\affiliation{RIKEN Center for Quantum Computing (RQC), Wako, Saitama, 351-0198, Japan}

\author{Hayato Goto}
\affiliation{RIKEN Center for Quantum Computing (RQC), Wako, Saitama, 351-0198, Japan}
\affiliation{Corporate Laboratory, Toshiba Corporation, Kawasaki, Kanagawa 212-8582, Japan}

\begin{abstract}
    To maximize the potential of quantum error-correcting codes, it is essential to develop high-performance decoders.
    The subsystem many-hypercube (MHC) codes have been developed to achieve both high encoding rates and low-weight syndrome-measurements, 
    but the introduction of gauge degrees of freedom makes decoding more challenging.
    In this work, we develop neural-network-based decoders for the subsystem MHC codes in a circuit-level noise model.
    We demonstrate that even the gauge-measurement information can be utilized for decoding by carefully arranging the syndrome-measurement sequence, improving the decoding performance.
    We further show that recurrent neural decoders outperform simple fully connected neural decoders, and can decode syndrome-measurement sequences longer than those used during training.
\end{abstract}

\maketitle


\section{Introduction}

Quantum error-correcting codes (QECCs) \cite{GottesmanWeb,Nielsen_Chuang_2010} provide a route toward large-scale fault-tolerant quantum computation, which is a critical challenge in modern quantum information science.
The surface codes \cite{KITAEV20032,bravyi1998quantumcodeslatticeboundary,PhysRevA.86.032324} are the current standard QECCs, enabling syndrome extraction through local stabilizer measurements on a two-dimensional grid of physical qubits and logical gate operations through local code-boundary manipulations \cite{Horsman_2012,Litinski2019gameofsurfacecodes}. 
However, advances in neutral-atom \cite{Bluvstein2024,Chiu2025,Bluvstein2026} and trapped-ion \cite{PhysRevLett.74.4091,Kielpinski2002,Ransford2026} quantum computers have lifted the locality restriction and shed light on high encoding-rate codes that require nonlocal connectivity.
Examples include quantum low-density parity-check (LDPC) codes \cite{
7456305,
6671468,
9490244,
kaufman2020newcosystolicexpanderstensors,
Panteleev2021degeneratequantum,
10.1145/3519935.3520017,leverrier2022quantumtannercodes,
Bravyi2024,
kasai2026breakingorthogonalitybarrierquantum}
and concatenated codes \cite{Yamasaki2024,doi:10.1126/sciadv.adp6388,Yoshida2025,xbzn-vn37,goto2026optimizedmanyhypercubecodeslower}.

In \cite{xbzn-vn37}, the authors have introduced subsystem codes based on the many-hypercube (MHC) codes \cite{doi:10.1126/sciadv.adp6388}, 
which are concatenated codes of the $[\![6,4,2]\!]$ and/or $[\![4,2,2]\!]$ quantum error-detecting codes.
Subsystem codes are defined by assigning some qubits as gauge degrees of freedom \cite{PhysRevLett.94.180501,PhysRevLett.95.230504,PhysRevA.73.012340}.
The introduction of gauge operators reduces the weights of the check operators since the gauge operators are irrelevant to the error correction and hence we can freely multiply them by any operators.
The subsystem MHC codes have equal syndrome-measurement weights at any levels of concatenation, in contrast to level-dependent weights in the original MHC codes. 

On the other hand, the introduction of gauge degrees of freedom increases error degeneracy, making decoding more challenging.
The number of degenerate error configurations grows double exponentially with the concatenation level, since the number of the gauge qubits increases exponentially.
Even for the bitflip noise model on the data qubits and for a single round of syndrome measurements, optimal decoding is computationally feasible only up to level 3 \cite{xbzn-vn37}.
This problem has been addressed using simple, fully connected neural-network decoders applicable to the code-capacity noise model, which outperform the belief propagation with ordered statistics decoding (BP-OSD) \cite{Panteleev2021degeneratequantum} up to level 4.
Based on this work, we here consider the level-2 and -3 subsystem MHC codes in a circuit-level noise model and develop neural-network-based decoders that maintain their decoding performance for larger numbers of syndrome-measurement rounds than the code distances.

So far, variety of neural decoders have been developed, such as those based on
multilayer perceptrons (MLPs)
\cite{
    Krastanov2017,
    Varsamopoulos_2018,
    PhysRevA.99.052351,
    PhysRevA.102.042411,
    8880492,
    9772289,
    Locher2023quantumerror,
    PhysRevResearch.6.L032004,
    xbzn-vn37
},
Boltzmann machines
\cite{
    PhysRevLett.119.030501
},
reinforcement learning
\cite{
    DOMINGOCOLOMER2020126353,
    PhysRevResearch.2.023230,
    Sweke_2021,
    matekole2022decodingsurfacecodesdeep
},
long short-term memory (LSTM)
\cite{
    Baireuther2018machinelearning, 
    Baireuther_2019,
    PhysRevResearch.7.013029
},
convolutional neural networks (CNNs)
\cite{
    Chamberland_2018,
    Breuckmann2018scalableneural,
    PhysRevResearch.2.033399,
    PhysRevLett.128.080505,
    Bordoni2023,
    zhang2023scalablefastprogrammableneural,
    Gicev2023scalablefast,
    Chamberland_2023,
    gu2026scalableneuraldecoderspractical
},
graph neural networks (GNNs)
\cite{
    Chamberland_2018,
    Breuckmann2018scalableneural,
    PhysRevResearch.2.033399,
    Ni2020neuralnetwork,
    PhysRevLett.128.080505,
    Bordoni2023,
    zhang2023scalablefastprogrammableneural,
    Gicev2023scalablefast,
    Chamberland_2023,
    gu2026scalableneuraldecoderspractical
},
Transformer
\cite{
    cao2023qecgptdecodingquantumerrorcorrecting,
    wang2023transformerqecquantumerrorcorrection,
    Choukroun2023,
    Bausch2024, 
    blue2026machinelearningdecodingcircuitlevel,
    ataides2025neuraldecodersuniversalquantum
}, and
Mamba
\cite{
    lee2025scalableneuraldecoderspractical
}.
The advantages of the neural decoders include the fact that the decoding time is constant and short, and that they don't need to be aware of the noise model, such as the code-capacity, phenomenological, or circuit-level noise models, and experimental data.
However, in order to strike a balance between training time, neural-network size, and the performance, it is necessary to select an appropriate neural network for a code.
In particular, neural decoders showing better performance than the standard decoders have been reported, e.g., for the surface codes \cite{Bausch2024} and quantum LDPC codes \cite{gu2026scalableneuraldecoderspractical}.

In this work, we develop two types of neural decoders for the $[\![4,2,2]\!]$ subsystem MHC codes in a circuit-level noise model, the MLP decoders and the LSTM decoders.
We use the MLP decoders to compare and optimize syndrome-measurement sequences. 
The flexibility of arranging the syndrome-measurement sequence is resulting from symmetry with respect to the coordinate exchange in the subsystem MHC codes.
We use the LSTM decoders to evaluate the generalization of the neural decoders, that is, the ability to decode larger numbers of measurement rounds than training ones.
The advantages of the MLP and LSTM lie in the fact that they do not require spatial structure unlike CNNs and that they are computationally lighter than Transformers.

This paper is organized as follows.
In Sec.~\ref{sec:smhc}, we first review the subsystem MHC codes, and then explain how the detectors are constructed.
In Sec.~\ref{sec:mlp}, we optimize the measurement sequence using the MLP decoders, and demonstrate that postselection improves decoding performance.
In Sec.~\ref{sec:lstm}, we show the generalization of the LSTM decoders. 
Finally, we conclude in Sec.~\ref{sec:conclusion}.

\section{Subsystem MHC codes}
\label{sec:smhc}

We first review the $[\![4,2,2]\!]$ subsystem MHC codes \cite{xbzn-vn37}.
We then explain the relation between the instantaneous stabilizer group (ISG) and a set of check measurements,
and also explain the detectors used as inputs into the neural decoders.

\subsection{Code definition}

The original $[\![4,2,2]\!]$ MHC codes \cite{doi:10.1126/sciadv.adp6388,goto2026optimizedmanyhypercubecodeslower} are defined by the code concatenation of the $[\![4,2,2]\!]$ quantum error-detecting codes.
The $[\![4,2,2]\!]$ quantum error-detecting code has stabilizers
\begin{align}
    SZ=Z_1Z_2Z_3Z_4,\quad SX=X_1X_2X_3X_4,
    \label{eq:level1mhcstabilizers}
\end{align}
and two sets of the logical operators
\begin{align}
    &Z^L_{1}=Z_1Z_2,\quad\, Z^L_{2}=Z_2Z_3,\\
    &X^L_{1}=X_2X_3,\quad X^L_{2}=X_1X_2,
\end{align}
where $Z_x$ and $X_x$ are the $x$-th qubit's operators at $x\in\{1,2,3,4\}$ on the $x$ axis.
In the following, we use $x,y\in\{1,2,3,4\}$ as the spatial coordinates and $i,j\in\{1,2\}$ as the indices of the operators.

The level-2 $[\![4,2,2]\!]$ MHC code is defined on 4 sets of the $[\![4,2,2]\!]$ codes, which are specified by the position $y\in\{1,2,3,4\}$ on the $y$ axis.
Based on the level-1 logical operators, the level-2 stabilizers are weight-8 operators given by
\begin{align}
    \begin{split}
        &SZ_{i}=Z^L_{i,1}Z^L_{i,2}Z^L_{i,3}Z^L_{i,4}, \\
        &SX_{i}=X^L_{(3-i),1}X^L_{(3-i),2}X^L_{(3-i),3}X^L_{(3-i),4},
    \end{split}
    \label{eq:level2mhcstabilizers}
\end{align}
where $i=1,2$ and $Z^L_{i,y}, X^L_{i,y}$ are the $i$-th logical operators at $y$.
The level-2 logical operators are weight-4 operators given by
\begin{align}
    Z^L_{ij}=Z^L_{i,j}Z^L_{i,(j+1)},\quad X^L_{ij}=X^L_{i,(3-j)}X^L_{i,(4-j)},
\end{align}
where $i,j=1,2$.
The stabilizer group of the level-2 MHC code is generated by four sets of the level-1 stabilizers in Eq.~(\ref{eq:level1mhcstabilizers}) and the level-2 ones in Eq.~(\ref{eq:level2mhcstabilizers}), for a total of 12 operators.
These stabilizer generators are the check operators to be measured to give the syndrome-measurement outcomes.
By repeating this concatenation, the level-$l$ $[\![4,2,2]\!]$ MHC code is constructed on an $l$-dimensional hyperlattice, and is a $[\![4^l,2^l,2^l]\!]$ code.

\begin{figure}
    \centering
    \includegraphics[width=0.47\textwidth]{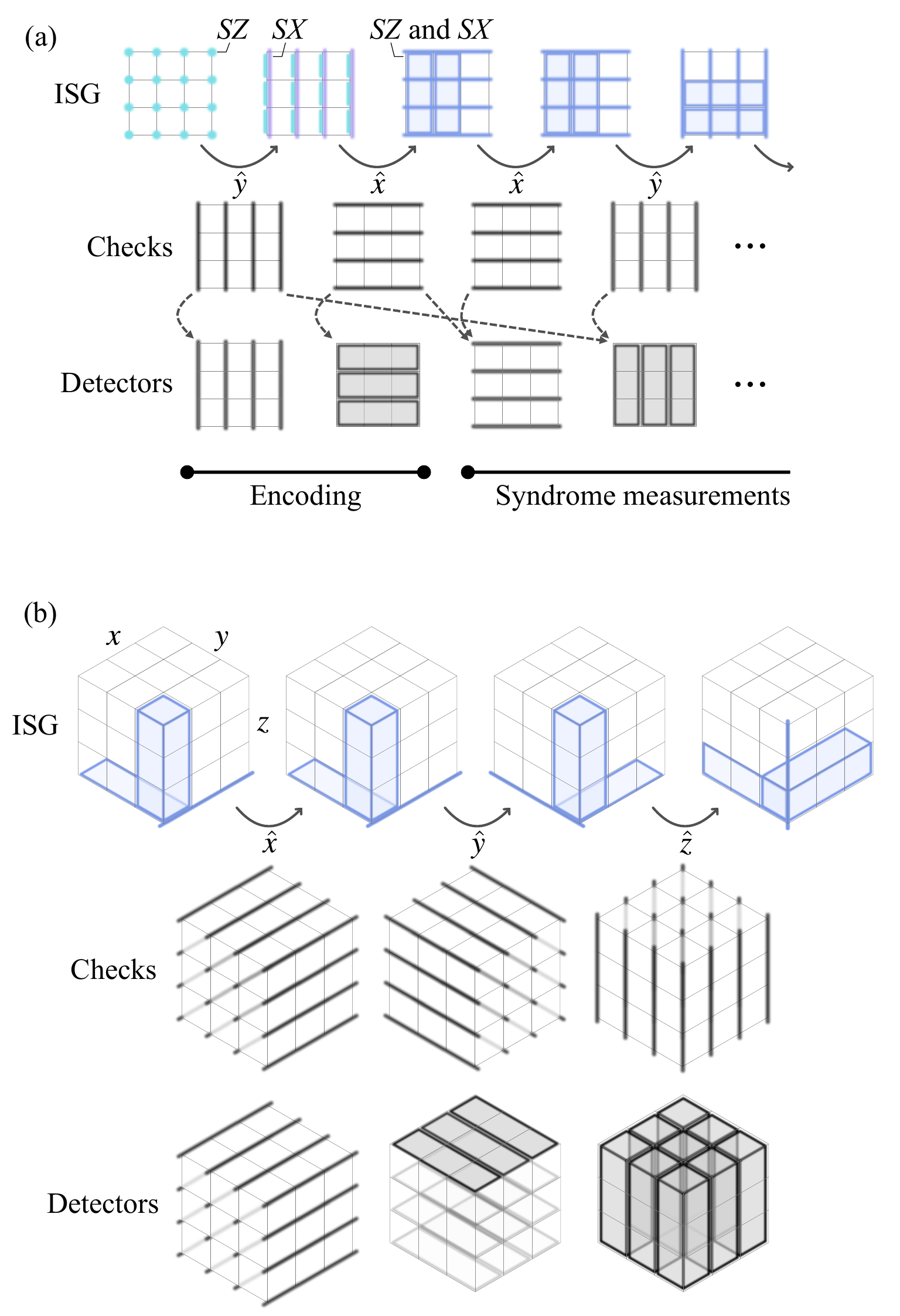}
    \caption{(a) The instantaneous stabilizer groups (ISGs) in the level-2 code are shown with the corresponding measured check operators and detectors.
    The first two sets of the measurements are used to encode the logical-$|0\rangle$ state and the subsequent sets are used for syndrome measurements.
    The figure corresponds to the measurement sequence $\hat{y}\hat{x}\hat{x}\hat{y}\cdots$. 
    The arrows indicate the flow that the detectors are defined from the measurement outcomes of the check operators.
    Only the detectors from the $Z$ check measurements are shown.
    (b) The ISGs in the level-3 code after the logical-$|0\rangle$ state encoding are shown for the measurement sequence $\hat{x}\hat{y}\hat{z}$.
    Only some of the ISG generators are shown while the others are their translations in two directions.
    \label{fig:isg}
    }
\end{figure}

The level-$l$ $[\![4,2,2]\!]$ subsystem MHC code is built from the MHC code with the same level by introducing weight-4 gauge operators \cite{xbzn-vn37}.
Let the gauge operators, which are simultaneously the check operators, be defined by the products of $X$ or $Z$ operators of qubits aligned in a straight line of the hyperlattice.
We denote by $g^{\hat{x}_a}$ a set of gauge operators whose qubits are aligned along the $x_a$ axis.
Each $g^{\hat{x}_a}$ contains $2\cdot 4^{l-1}$ operators specified by the coordinate $(x_1,\cdots, x_{a-1}, x_{a+1},\cdots, x_l)$ in the $(l-1)$-dimensional hyperplane perpendicular to the $x_a$ axis.
Specifically, the gauge operators of the level-2 $[\![4,2,2]\!]$ subsystem MHC code are weight-four operators given by [see Fig.~\ref{fig:isg} (a)]
\begin{align}
    &g^{Z,\hat{x}}_{y}=Z_{1y}Z_{2y}Z_{3y}Z_{4y},\quad\, g^{Z,\hat{y}}_{x}=Z_{x1}Z_{x2}Z_{x3}Z_{x4}, \\
    &g^{X,\hat{x}}_{y}=X_{1y}X_{2y}X_{3y}X_{4y},\quad g^{X,\hat{y}}_{x}=X_{x1}X_{x2}X_{x3}X_{x4},
\end{align}
where $X_{xy}, Z_{xy}$ are operators of a qubit at $x,y\in\{1,2,3,4\}$.
Here, the operators with $\hat{x}$ ($\hat{y}$) on the superscript are elements of $g^{\hat{x}}$ ($g^{\hat{y}}$).

The gauge operators form a gauge group $\mathcal{G}$ \cite{PhysRevLett.94.180501,PhysRevLett.95.230504,PhysRevA.73.012340}.
The stabilizer group is the center of $\mathcal{G}$, that is, $\mathcal{S}=\mathcal{Z}(\mathcal{G})\cap\mathcal{G}$. 
The stabilizer generators of the level-2 $[\![4,2,2]\!]$ subsystem MHC code consist of (\ref{eq:level2mhcstabilizers}) and products of the level-1 stabilizers (\ref{eq:level1mhcstabilizers}) given by $SZ_ySZ_{y+1}$ and $SX_ySX_{y+1}$ for $y\in[1,3]$.
There are ten stabilizer generators of weight 8.
The logical operators are the same as those of the MHC codes.
In the following, we refer to the level-$l$ subsystem MHC code as the level-$l$ code for simplicity.
Notice that the gauge structure of the subsystem MHC codes has close similarity with the Blocklet concatenation in \cite{litinski2025blockletconcatenationlowoverheadfaulttolerant} at least up to level 3.

The gauge operators are measured using ancilla qubits.
We use a syndrome extraction method developed in \cite{Reichardt_2021}, which can detect harmful errors that spread into the data qubits (see the circuit in \cite{xbzn-vn37}).
We need $2\cdot4^{l-1}$ ancilla qubits to measure all the $X$ and $Z$ check operators in $g^{\hat{x}_a}$. 
The ancilla qubits can be reused in measuring a different set of the check operators.

\subsection{Instantaneous stabilizer groups}

In circuit-level noise models, we measure the check operators multiple times to infer the most probable errors and preserve quantum information stored in physical qubits.
The check measurements have some flexibility regarding their order and sequence.
Since the check operators in the same group $g^{\hat{x}_a}$ do not have overlaps, we regard the measurements of the check operators in $g^{\hat{x}_a}$ as the minimum unit of the syndrome measurements.
At level-$l$, there are $l$ sets of the check operators $\{g^{\hat{x}_1},g^{\hat{x}_2},\cdots,g^{\hat{x}_l}\}$, the measurements of which are denoted by $\hat{x}_a\,(a\in[1,l])$ (see checks in Fig.~\ref{fig:isg}).
The syndrome-measurement sequence is specified by $\hat{x}_{a(1)}\hat{x}_{a(2)}\cdots$, where $a(t)$ denotes the direction of the $t$-th check measurements.
When we repeat $n$ times the same sequence $\hat{x}_a\cdots \hat{x}_b$, we use an abbreviated notation $(\hat{x}_a\cdots \hat{x}_b)^n$.

Every time a set of check operators is measured, the quantum state can be modified while the quantum information is preserved since all the check operators commute with all the logical operators.
This indicates that the ISG that specifies the code space is updated by the check operators that are just measured.
Let the ISG before the $t$-th check measurements be denoted by $\mathcal{S}(t)$.
After we measure $\hat{x}_{a(t)}$, the ISG becomes $\mathcal{S}(t+1)$, which contains the check operators in $g^{\hat{x}_{a(t)}}$ while other operators in $\mathcal{S}(t)$ may be reconfigured so that they commute with $g^{\hat{x}_{a(t)}}$ (Fig.~\ref{fig:isg}).
In a special case, the ISG is unchanged when the same check operators are measured again.
Notice that the ISG is not the same as the stabilizer group $\mathcal{S}$ that are defined by the gauge group $\mathcal{G}$.
$\mathcal{S}$ is in general the intersection of $\mathcal{S}(1),\mathcal{S}(2),\cdots$, that is, $\mathcal{S}$ is always a subgroup of $\mathcal{S}(t)$.  

Before the syndrome measurements, we initialize the data qubits by encoding the logical-$|0\rangle$ state.
That is, we start with all data qubits in the $|0\rangle$ state and measure all the check operators once.
In particular, to prepare the logical-$|0\rangle$ state of the original MHC codes, we measure in order of $\hat{x}_l\hat{x}_{l-1}\cdots\hat{x}_1$ [Fig.~\ref{fig:isg} (a)].
After the measurements, the ISG is the same as the stabilizer group of the original MHC codes.

For simplicity, we assume that the total number of the check measurements is the same for all the directions, since the subsystem MHC is symmetric under permutations of the coordinates.
We regard the number of the same check measurements as the number of rounds.
For instance, at level 3, syndrome measurements of both $\hat{x}\hat{y}\hat{z}\hat{x}\hat{y}\hat{z}$ and $\hat{x}\hat{x}\hat{y}\hat{y}\hat{z}\hat{z}$ correspond to 2 rounds.

\subsection{Detectors}

Measurement outcomes are arranged to form detectors.
By definition, the detectors are deterministic combinations of the measurement outcomes when errors are absent \cite{Gidney2021stimfaststabilizer}.
In particular, we define detectors from the outcomes of the check measurements $\hat{x}_{a(t)}$ by the difference of such combinations between $t$ and the time when the same check operators are last measured.
The definition of the detectors is uniquely determined by the syndrome-measurement sequence.

Specifically, suppose we measure the same set of the check operators twice, like $\cdots\hat{x}_a\hat{x}_a$ without errors.
While the outcomes of the first measurements may or may not be deterministic depending on the initial state, the difference between the first and second measurements is deterministic since the ISG after the first measurements contain the corresponding set of the check operators $g^{\hat{x}_a}$.
There are $2\cdot 4^{l-1}$ detectors from the second measurements.
This indicates that the measurement outcomes not just of the stabilizers in $\mathcal{S}$ but also of the gauge operators can be used to infer errors.

On the other hand, if we measure some other sets of the check operators between $\hat{x}_a$ like $\cdots\hat{x}_a\hat{x}_b\hat{x}_c\hat{x}_a$, the outcomes of the second measurements no longer correspond one-to-one to the detectors.
The combinations of the measurement outcomes that comprise the detectors are determined by the directions $x_b$ and $x_c$.
When we measure $\hat{x}_b$ after the first $\hat{x}_a$ measurements, the ISG contains products of two check operators in $g^{\hat{x}_a}$ separated by $e_b$, which is the unit vector along the $x_b$ direction [Fig.~\ref{fig:isg} (a)].
These operators have weight $8$.
If we further measure $\hat{x}_c$, the ISG contains products of two weight-8 operators in the previous ISG separated by $e_c$ [see the stabilizer generators along $x$ in Fig.~\ref{fig:isg} (b)].
The deterministic outcomes from the second $\hat{x}_a$ measurements are the eigenvalues of the weight-16 operators in the intersection of $g^{\hat{x}_a}$ and the ISG just before the measurements, compared with those from the first measurements.
Thus, if we measure $n$ different directions of the check operators between two $\hat{x}_a$'s, there appear $2\cdot4^{l-1-n}3^n$ detectors whose operator weight is $4\cdot 2^n$.
Specifically, if we measure $\hat{z}\hat{y}\hat{x}\hat{x}\hat{y}\hat{z}$ in the level-3 code, there are, respectively, $32,24,18$ detectors after the second $\hat{x}$ measurements, corresponding operators of which coincide with the stabilizer generators of the original level-3 MHC code [Fig.~\ref{fig:isg} (b)].
Thus, this measurement sequence maximally utilizes the gauge information.
On the other hand, if we measure $\hat{x}\hat{y}\hat{z}\hat{x}\hat{y}\hat{z}$, the numbers of detectors after the second $\hat{x}$ measurements are all 18, where we cannot use the gauge information at all.

We can increase the number of detectors by repeatedly measuring the same check operators many times, which reduces the measurement errors.
However, this leads to a long interval between the check measurements of the other directions, resulting in less information for finding errors.
Thus, we need to balance the two effects by optimizing the measurement sequence, which will be discussed in the next section.
Notice that, for subsystem Calderbank–Shor–Steane (CSS) codes, a similar idea has been studied in \cite{PhysRevX.11.031039}, where $Z$ and $X$ check operators are each measured repeatedly to increase the number of detectors.

For practical purpose, we measure all the data qubits after some rounds of the syndrome measurements.
This process is necessary to identify the total logical errors and configure the last set of detectors, of which there are $2(4^l-3^l)$.

In Fig.~\ref{fig:isg}, the detectors and the change of the ISG for the level-2 and -3 codes are shown for particular sequences of the syndrome measurements.
Notice that while in ISG we can omit some of the generators that are redundant, such operators need to be included in the definition of the detectors since they provide independent information for decoding.
Specifically, in Fig.~\ref{fig:isg} (a), the ISGs after the logical-$|0\rangle$ state encoding are missing weight-8 rectangular operators ($SZ$ and $SX$) on the far right or at the topmost, while such operators are included in the detectors after the final $\hat{y}$ measurements.

\section{MLP decoders}
\label{sec:mlp}

Here, we optimize the measurement sequence of the subsystem MHC codes by using MLP as the neural-network structure of the decoder.
Throughout this paper, we assume a circuit-level noise model that includes single-qubit and two-qubit depolarizing errors after Clifford operators, and bit-flip errors before ancilla measurements and after resets of the ancilla qubits with the same physical error rate $p$.
Hereafter, we focus on the level-2 and -3 codes.

Since we use the logical-$|0\rangle$ state encoding, we infer $X$ errors, which flip the eigenvalues of the logical-$Z$ operators.
That is, we consider the detector outcomes only of $Z$ check operators and the eigenvalues of the logical-$Z$ operators after the syndrome measurements.
We use the former data as the inputs for the MLP and the latter as the labels.
On the other hand, $Z$ errors can also be inferred similarly by initializing all data qubits in the $|+\rangle$ state and by the logical$-|+\rangle$ state encoding.
The detector outcomes and the total logical errors are simulated by \textit{Stim} \cite{Gidney2021stimfaststabilizer}.
MLPs output a set of $2^l$ real values between 0 and 1, which give the predictions by the closest integers.
If all the $2^l$ predictions match the labels, the decoding is successful.

We use MLPs consisting of three hidden layers each with 256 (1024) nodes for the level-2 (-3) code,
and train them using batch size 64 (256) and the training data generated at physical error rate $p=0.01$ (0.002).
The learning rate is annealed from $10^{-3}$ by the cosine annealing down to $10^{-6}$.
We use ReLU as the activation function, the binary cross entropy as a loss function, and Adam with decoupled weight decay as an optimizer.
The outputs of the MLP are logits $\in(-\infty,\infty)^{2^l}$, which are transformed to the predictions by the Sigmoid function.
Throughout this section, we consider six rounds of the syndrome measurements.

\subsection{Measurement sequence}
\label{sec:measurementorder}

\begin{figure}
    \includegraphics[width=0.43\textwidth]{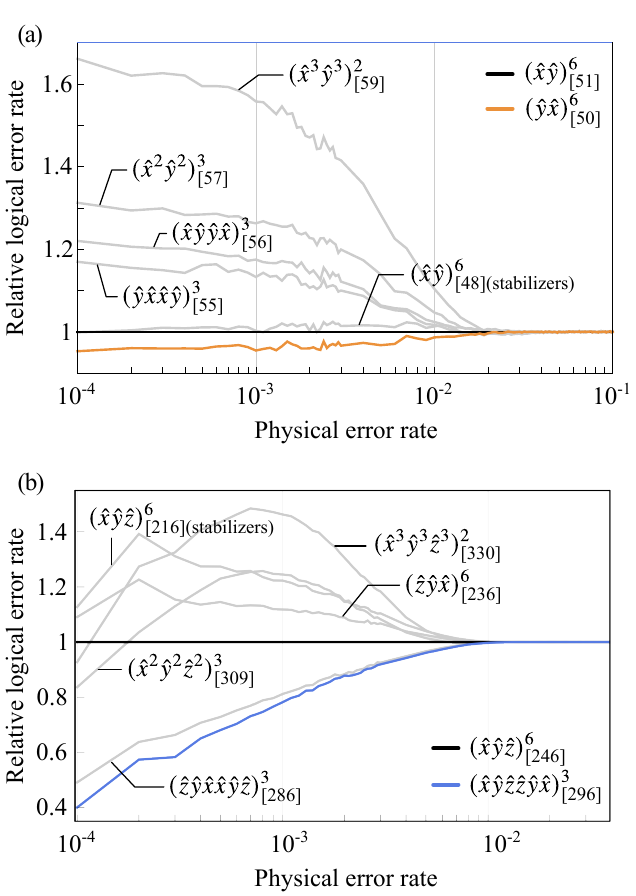}
    \caption{
        Relative logical error rates per round of the MLP decoders compared with (a) $(\hat{x}\hat{y})^6$ for the level-2 code, and (b) with $(\hat{x}\hat{y}\hat{z})^6$ for the level-3 code.
        The measurement sequences have a common measurement rounds of six. The numbers in the subscript with bracket represent the total number of detectors.
        The best performance results are highlighted in yellow and blue, respectively.
        \label{fig:mlp}
    }
\end{figure}

First, we optimize the measurement sequence by comparing the performance over a total of six rounds.
The logical error rate per round of the MLP decoders as a function of the physical error rate is shown in Fig.~\ref{fig:mlp}.
The measurement sequences do not include the logical-$|0\rangle$ state encoding.
We consider repeated measurements of the same set of the check operators up to three times, e.g., $(\hat{x}^3\hat{y}^3)^2$.
Numbers in the subscript with the square bracket is the total number of detectors including those in the logical-$|0\rangle$ state encoding, syndrome measurements, and the final data-qubit measurements.
That is, they are the input dimension of the MLP.
Data with ``(stabilizers)'' indicates that only the eigenvalues of the stabilizer $\mathcal{S}$ are used for decoding without the additional gauge information.
The logical error rate is shown in comparison with those of $(\hat{x}\hat{y})^6$ and $(\hat{x}\hat{y}\hat{z})^6$, respectively, and the best performance results are highlighted by the yellow and blue lines.

At level 2, the measurement order of $(\hat{y}\hat{x})^6$ shows the best performance.
This indicates that measuring all the directions equally frequently is more important than measuring the same direction repeatedly.
In addition, $(\hat{y}\hat{x})^6$ performs slightly better than $(\hat{x}\hat{y})^6$ since we measure $\hat{y}\hat{x}$ in advance in the logical-$|0\rangle$ state encoding, that is, we measure totally $(\hat{y}\hat{x})^7$ in the best performance case.
On the other hand, whether we include the gauge information in decoding is less significant.

The situation is slightly different at level 3, where $(\hat{x}\hat{y}\hat{z}\hat{z}\hat{y}\hat{x})^3$ performs the best.
This sequence balances between the measurement frequency and the number of the detectors.
Particularly, the number of detectors per round is increased to the same number as those of the stabilizer generators of the original level-3 MHC code.
Moreover, an effective use of the gauge information clearly improves the decoding performance.

In the rest of this paper, we use measurement sequences $(\hat{y}\hat{x})^r$ and $(\hat{x}\hat{y}\hat{z}\hat{z}\hat{y}\hat{x})^{r/2}$ for $r$ rounds of syndrome measurements in the level-2 and level-3 codes, respectively.

\subsection{Postselection}

The MLP decoders output $2^l$ real-number predictions between 0 and 1.
When all the predictions are close to either 0 or 1, the decoders are confident in the prediction, while they are less confident when some are close to 0.5.
In general, the decoders are less confident in their predictions for errors that occur less frequently in training.
Since rare errors are likely to have high error weights that are not recoverable, discarding such cases can improve the decoding performance.

\begin{figure}
    \includegraphics[width=0.43\textwidth]{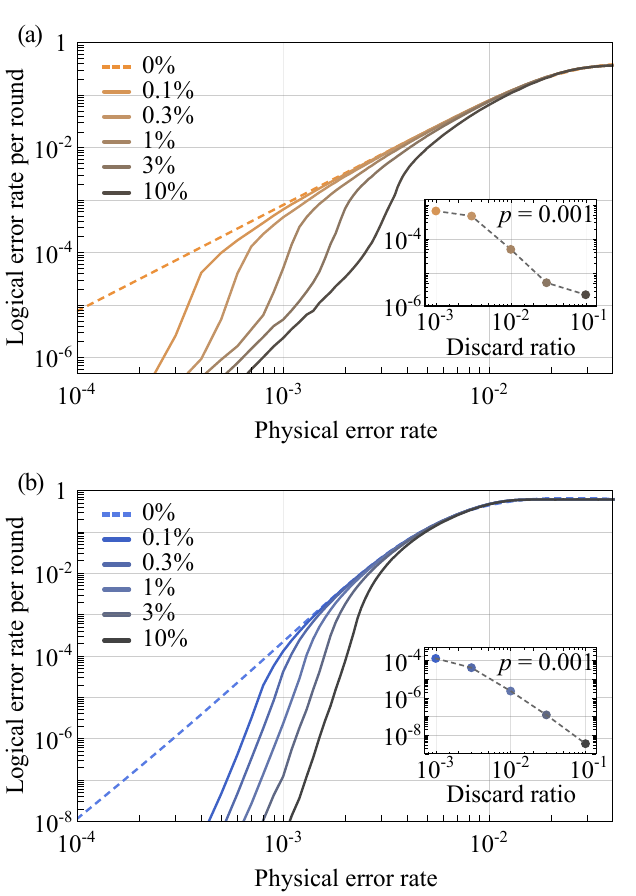}
    \caption{
        Logical error rates per round after the postselection based on the mean entropy of the decoder's predictions for (a) the level-2 and (b) level-3 codes.
        The same decoders as Fig.~\ref{fig:mlp} are used.
        The percentage shown in the plot indicates the proportion of the discarded samples that have larger mean entropy (less confident).
        The dashed lines are the performance without postselection.
        The insets show the performance at physical error rate $p=0.001$ as a function of the discard ratio.
        \label{fig:postselection}
    }
\end{figure}

To do so, we postselect the decoding results based on the entropy of the output.
When the output of the MLP decoder is $(p_1,\cdots,p_{2^l})\, (p_j\in[0,1])$, the mean Shannon entropy over the $2^l$ outputs is defined by
\begin{align}
    s=-\frac{1}{2^l}\sum_{j=1}^{2^l}\left[p_j\log p_j+(1-p_j)\log(1-p_j)\right],
\end{align}
which is close to 0 when the prediction is reliable.
Figure ~\ref{fig:postselection} shows the logical error rate per round after the postselection, where we discard samples with higher mean entropy.
For comparison, we also plot the performance without it from Sec.~\ref{sec:measurementorder} by the dashed lines.
At level 3, the logical error rate at physical error rate $p=0.001$ can be improved by four orders of magnitude by discarding 10\% samples. 
We confirmed that whether we use the maximum or mean entropy over the $2^l$ outputs does not affect the logical error rate.

\section{LSTM decoders}
\label{sec:lstm}

\begin{figure}
    \includegraphics[width=0.44\textwidth]{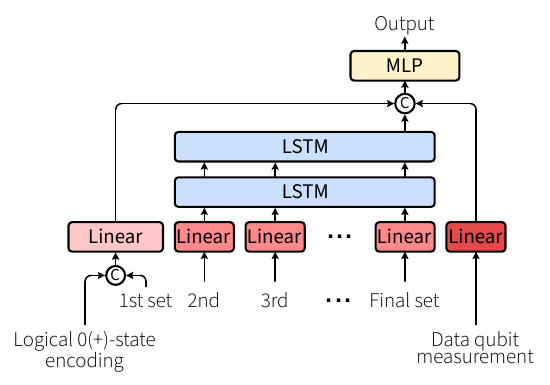}
    \caption{
        The neural-network architecture of the LSTM decoders.
        The embedding layer consists of three types of linear mappings followed by ReLU.
        The outputs from the syndrome-measurements except the first are processed by two LSTM layers and then fed into the MLP layer after concatenated with the first and last outputs of the embedding layer.
        Circles with C represent the concatenation.
        \label{fig:lstmarchitecture}
    }
\end{figure}

In the last section, we presented the MLP decoders trained and evaluated with six-round data.
For decoding longer sequences, we develop neural decoders that incorporate LSTM \cite{Baireuther2018machinelearning,Baireuther_2019,PhysRevResearch.7.013029}, which we refer to as the LSTM decoders.
The architecture of the LSTM decoders is shown in Fig.~\ref{fig:lstmarchitecture}.
The input data consists of the detectors from the logical-$|0\rangle$ state encoding, $r$ ($r/2$) sets of the syndrome measurements for the level-2 (-3) code, and the data-qubit measurement.
Notice that for the level-3 code, a single set $\hat{x}\hat{y}\hat{z}\hat{z}\hat{y}\hat{x}$ consists of two rounds, and thus the number of sets is half of that of the measurement rounds.
First, the input data is decomposed into $r+1$ ($r/2+1$) feature vectors, the first vector of which is the detectors before the second set of the syndrome measurement.
The feature vector from the first set is separated from those after the second set, since it may have different dimensions (though it turned out not to be necessary after the optimization in Sec.~\ref{sec:measurementorder}).
The feature vectors are mapped to higher-dimensional representation by the embedding layer consisting of three types of linear layers followed by layer normalization and ReLU.
We set the embedding dimension $d$ common to all the embeddings.
The embeddings from repeated syndrome measurements are fed into the recurrent part, which shares the same hidden dimension $d$.

The recurrent part consists of two layers of the LSTM.
Each LSTM layer stores short- and long-term memories.
The first LSTM layer updates its long-term memory by processing the short-term memory and an embedding input, 
and then outputs the short-term memory for the next input and the input for the second LSTM layer.
We use the output from the second LSTM layer after processing the last output from the first LSTM layer.
The output from the second LSTM is concatenated with the other embeddings, and then fed into the MLP, which consists of three hidden layers with nodes $3d$.
We predict the logical errors from the output of the MLP layer.

The numbers of layers in LSTM and MLP are determined so that increasing them does not improve the performance within a fixed training time.
We tune the embedding dimensions, the physical error rate of the training data, and the learning rate.

\subsection{Results}

For the level-2 code, we use embedding dimension $d=256$, batch size 64, and learning rate $10^{-3}$.
The neural network is trained by syndrome outcomes generated randomly from 8 to 16 rounds with a training physical error rate of $p=0.01$.

\begin{figure}
    \includegraphics[width=0.44\textwidth]{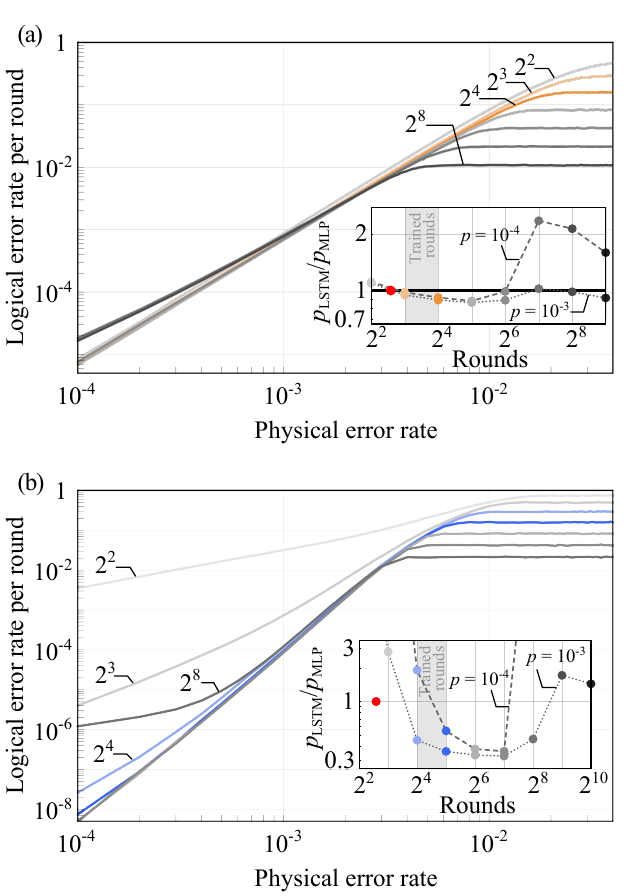}
    \caption{
        Performance of the LSTM decoders for the (a) level-2 and (b) -3 codes.
        The inset shows the relative logical error rate per round for physical error rates $p=10^{-3}$ and $10^{-4}$ compared with the MLP decoder's performance for six rounds.
        The shaded regions indicate the rounds that are used in training.
        The red points show the performance of the MLP decoders shown in Sec.~\ref{sec:mlp}.
        \label{fig:lstm}
    }
\end{figure}

The logical error rate per round evaluated from 4 to 256 rounds is shown in Fig.~\ref{fig:lstm} (a).
The result shows that the performance is generalized to sequences much longer than the code distance and the training rounds.
However, the performance in particular at smaller physical error rates ($p\sim 10^{-4}$) is not maintained when the number of rounds exceeds 64 [Fig.~\ref{fig:lstm} (a) inset].
Notice that at high physical error rates the logical error rate converges to $1-2^{-2^l}$, where the logical error is completely unrecoverable and thus the decoding is successful randomly with probability $1/2^{2^l}$.
In this case, the logical error rate per round converges to a plateau value $1-2^{-2^l/r}\simeq (2^l\ln 2)/r$, where $r$ is the number of measurement rounds.
For this reason, the logical error rate of $256$ rounds looks better than that of $128$ [see the inset of Fig.~\ref{fig:lstm} (a)].
The performance peaks at $32$ rounds.

For the level-3 code, we use embedding dimension $d=512$, batch size 256, and learning rate $10^{-3}$.
The neural network is trained using simulated data from 16 to 32 rounds with a training physical error rate of $p=0.002$.

\begin{figure}
    \includegraphics[width=0.44\textwidth]{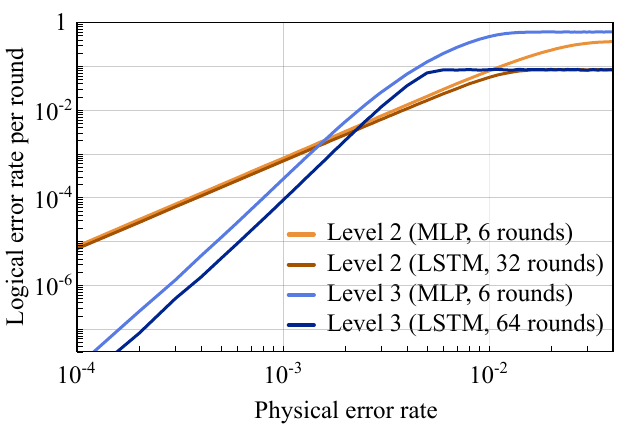}
    \caption{
        Best performance results of the MLP and LSTM decoders.
        \label{fig:bests}
    }
\end{figure}

Figure \ref{fig:lstm} (b) shows the logical error rate per round evaluated from 4 to 256 rounds. 
The performance is maintained at least up to four times longer (128 rounds) than the maximum of the training rounds, and is the best at 64 or 128 rounds.
This behavior stands in contrast to that of the MLP decoders where the performance gets worse monotonically as the number of the measurement rounds increases due to the increased input dimension.
In Fig.~\ref{fig:bests}, we summarize the best performance results of the MLP and LSTM decoders.
By using the LSTM decoder, the logical error rate is improved by three times smaller than the simple MLP decoder, in addition to the ability to decode more rounds.
The error threshold is estimated at $p\simeq 0.2\%$.

\section{Conclusion}
\label{sec:conclusion}

In this work, we developed neural-network-based decoders for the $[\![4,2,2]\!]$ subsystem MHC codes in a circuit-level noise model.
We developed MLP decoders for the purpose of the optimization of the syndrome-measurement sequence
and LSTM decoders for decoding more rounds of syndrome measurements.

The MLP decoders can process a fixed number of syndrome measurement rounds, thus we can compare the measurement sequence on the same footing.
The results indicate that the best performance is obtained by $(\hat{y}\hat{x})^r$ at level 2 and $(\hat{x}\hat{y}\hat{z}\hat{z}\hat{y}\hat{x})^{r/2}$ at level 3.
The measurement sequence is characterized by two factors.
One is how often the measurement outcomes for all the directions are available, 
and the other is how often the measurements for the same direction are repeated.
The former enables detecting errors from different types of information,
and the latter provides reliable information from a single direction due to the increased number of detectors and reduces measurement errors.
At level 2, the former is more favored, while at level 3, a sequence that balances the both is favored.
This indicates that at level 3, using the gauge information improves the decoding performance.
We also showed that the performance is improved by the orders of magnitude by the postselection based on the mean Shannon entropy of the predictions by the decoders.

The LSTM decoders can process arbitrary numbers of measurement rounds.
The neural networks consist of two LSTM layers followed by three layers of MLP.
Using the best performance sequence, 
we showed that the decoding performance is generalized to as many rounds at least 4 times as the maximum training rounds, which is much longer than the code distance.

As for future direction, it is intriguing to develop structure-aware (neural) decoders.
Since the subsystem MHC codes are symmetric under permutations of the coordinates, incorporating this symmetry into decoders could improve the decoding performance.
In particular, this direction of study is particularly successful in neural decoders for the quantum LDPC codes \cite{gu2026scalableneuraldecoderspractical}.

\begin{acknowledgements}
    This work was supported by JST Moonshot R\&D Grant Number JPMJMS256E.
\end{acknowledgements}


%

\end{document}